\documentclass{article}

\usepackage{amssymb}
\usepackage{graphics}
\usepackage{amsmath}

\begin{document}
\renewcommand{\Bbb}{\mathbb}

\thispagestyle{empty}

\newcommand{\al}{\alpha}
\newcommand{\bet}{\beta}
\newcommand{\ga}{\gamma}
\newcommand{\del}{\delta}
\newcommand{\ep}{\epsilon}
\newcommand{\epx}{\varepsilon}
\newcommand{\ze}{\zeta}
\renewcommand{\th}{\theta}
\newcommand{\thx}{\vartheta}
\newcommand{\io}{\iota}
\newcommand{\la}{\lambda}
\newcommand{\ka}{\kappa}
\newcommand{\pix}{\varpi}
\newcommand{\rhx}{\varrho}
\newcommand{\si}{\sigma}
\newcommand{\six}{\varsigma}
\newcommand{\yp}{\upsilon}
\newcommand{\om}{\omega}
\newcommand{\phx}{\varphi}
\newcommand{\Ga}{\Gamma}
\newcommand{\De}{\Delta}
\newcommand{\Th}{\Theta}
\newcommand{\La}{\Lambda}
\newcommand{\Si}{\Sigma}
\newcommand{\Yp}{\Upsilon}
\newcommand{\Om}{\Omega}

\renewcommand{\L}{\cal{L}}
\newcommand{\M}{\cal{M}}
\newcommand{\G}{{\cal G}}
\newcommand{\J}{{\cal J}}
\newcommand{\Jb}{\bar{\cal J}}
\newcommand{\be}{\begin{equation}}
\newcommand{\ee}{\end{equation}}
\newcommand{\bea}{\begin{eqnarray}}
\newcommand{\eea}{\end{eqnarray}}
\newcommand{\jt}{\tilde{J}}
\newcommand{\Ra}{\Rightarrow}
\newcommand{\lra}{\longrightarrow}
\newcommand{\ti}{\tilde}
\newcommand{\pj}{\prod J}
\newcommand{\pjt}{\prod\tilde{J}}
\newcommand{\prb}{\prod b}
\newcommand{\prc}{\prod c}
\newcommand{\bft}{|\tilde{\phi}>}
\newcommand{\bfj}{|\phi>}
\newcommand{\lan}{\langle}
\newcommand{\ran}{\rangle}
\newcommand{\bz}{\bar{z}}
\newcommand{\bJ}{\bar{J}}
\newcommand{\tsp}{\tau\hspace{-1mm}+\hspace{-1mm}\si}
\newcommand{\tsm}{\tau\hspace{-1mm}-\hspace{-1mm}\si}
\newcommand{\vacr}{|0\rangle}
\newcommand{\vacl}{\langle 0|}
\newcommand{\IFF}{\Longleftrightarrow}
\newcommand{\phr}{|phys\ran}
\newcommand{\phl}{\lan phys|}
\newcommand{\non}{\nonumber\\}
\newcommand{\tg}{\tilde{g}}
\newcommand{\tM}{\ti{M}}
\newcommand{\hd}{\hat{d}}
\newcommand{\hL}{\hat{L}}
\newcommand{\sir}{\si^\rho}
\newcommand{\mf}{\mathfrak}
\newcommand{\mfg}{{\mathfrak{g}}}
\newcommand{\mfbg}{{\bar{\mathfrak{g}}}}
\newcommand{\mfk}{{\mathfrak{k}}}
\newcommand{\mfc}{{\mathfrak{c}}}
\newcommand{\mfbc}{{\bar{\mathfrak{c}}}}
\newcommand{\mfbk}{{\bar{\mathfrak{k}}}}
\newcommand{\mfn}{{\mathfrak{n}}}
\newcommand{\mfbn}{{\bar{\mathfrak{n}}}}
\newcommand{\mfh}{{\mathfrak{h}}}
\newcommand{\mfbh}{{\bar{\mathfrak{h}}}}
\newcommand{\mfp}{{\mathfrak{p}}}
\newcommand{\mfbp}{{\bar{\mathfrak{p}}}}
\newcommand{\mfU}{{\mf U}}
\newcommand{\p}{\partial}
\newcommand{\pb}{\bar{\hspace{0.2mm}\partial}}
\newcommand{\pl}{\partial_{+}}
\newcommand{\pmi}{\partial_{\hspace{-0.5mm}-}}
\newcommand{\Lg}{\L^{(\mfg})}
\newcommand{\Lgk}{\L^{(\mfg,\mfk)}}
\newcommand{\Lgkp}{\L^{(\mfg,\mfk')}}
\newcommand{\Lk}{\L^{(\mfk)}}
\newcommand{\Lkp}{\L^{(\mfk')}}
\newcommand{\Lc}{\L^{(\mfc)}}
\newcommand{\Lgg}{\L^{(\mfg,\mfg)}(\la)}
\newcommand{\Lgp}{\L^{(\mfg,\mfg')}(\la)}
\newcommand{\Mgp}{\M^{(\mfg,\mfg')}(\la)}
\newcommand{\Mgk}{\M^{(\mfg,\mfk)}(\la)}
\newcommand{\Hgk}{H^{(\mfg,\mfk)}}
\newcommand{\Mgkp}{\M^{(\mfg,\mfk^\prime)}}
\newcommand{\Hgp}{H^{(\mfg,\mfg')}}
\newcommand{\vo}{v_{0\la}}
\newcommand{\Ml}{\M(\la)}
\newcommand{\Ll}{\L(\la)}
\newcommand{\RR}{\mathbb{R}}
\newcommand{\CC}{\mathbb{C}}
\newcommand{\sltwo}{\ensuremath{\mathfrak{sl}_{2}}}

\newcommand{\e}[1]{\label{e:#1}\end{eqnarray}}
\newcommand{\ind}{\indent}
\newcommand{\noi}{\noindent}
\newcommand{\np}{\newpage}
\newcommand{\hs}{\hspace*}
\newcommand{\vs}{\vspace*}
\newcommand{\nl}{\newline}
\newcommand{\bqu}{\begin{quotation}}
\newcommand{\equ}{\end{quotation}}
\newcommand{\bit}{\begin{itemize}}
\newcommand{\eit}{\end{itemize}}
\newcommand{\ben}{\begin{enumerate}}
\newcommand{\een}{\end{enumerate}}
\newcommand{\ba}{\begin{array}}
\newcommand{\ea}{\end{array}}
\newcommand{\ul}{\underline}
\newcommand{\nn}{\nonumber}
\newcommand{\lef}{\left}
\newcommand{\rig}{\right}
\newcommand{\fra}{\twelvefrakh}
\newcommand{\Bb}{\twelvemsb}
\newcommand{\bT}{\bar{T}(\bz)}
\newcommand{\dagg}{^{\dagger}}
\newcommand{\qd}{\dot{q}}
\newcommand{\cP}{{\cal P}}
\newcommand{\hg}{\hat{g}}
\newcommand{\hh}{\hat{h}}
\newcommand{\hpg}{\hat{g}^\prime}
\newcommand{\htg}{\tilde{\hat{g}}^\prime}
\newcommand{\pri}{\prime}
\newcommand{\bis}{{\prime\prime}}
\newcommand{\lap}{\la^\prime}
\newcommand{\rhop}{\rho^\prime}
\newcommand{\Dgp}{\Delta_{g^\prime}^+}
\newcommand{\Dg}{\Delta_g^+}
\newcommand{\Pro}{\prod_{n=1}^\infty (1-q^n)}
\newcommand{\Pg}{P^+_{\hg}}
\newcommand{\Pgp}{P^+_{\hg\pri}}
\newcommand{\hmu}{\hat{\mu}}
\newcommand{\hnu}{\hat{\nu}}
\newcommand{\hrho}{\hat{\rho}}
\newcommand{\gp}{g^\prime}
\newcommand{\pp}{\prime\prime}
\newcommand{\CM}{\hat{C}(g',M')}
\newcommand{\CI}{\hat{C}(g',M^{\prime (1)})}
\newcommand{\CL}{\hat{C}(g',L')}
\newcommand{\HL}{\hat{H}^p (g',L')}
\newcommand{\HMI}{\hat{H}^{p+1}(g',M^{\prime (1)})}
\newcommand{\da}{\dagger}
\newcommand{\asu}{\ensuremath{\widehat{\hspace{0.5mm}\mathfrak{su}}(1,1)\,}}

\newcommand{\wro}{w^\rho}
\renewcommand{\Box}{\rule{2mm}{2mm}}
\begin{flushright}
July 10, 2002 \\
\end{flushright}
\vs{10mm}

\begin{center}

{\Large{\bf CFT description of three-dimensional Kerr-de Sitter spacetime}}
\\
\vspace{10 mm}
{\large{Jens Fjelstad$^{*}$ \footnote{email: jens.fjelstad@kau.se}\\
Stephen Hwang$^{*}$ \footnote{email: stephen.hwang@kau.se}\\
Teresia M{\aa}nsson$^{\dagger}$ \footnote{email: teresia@physto.se}}
\vspace{4mm}\\
$^{*}$Department of Physics,\\ Karlstad
University, SE-651 88 Karlstad, Sweden\\
$^{\dagger}$Institute of Theoretical Physics,\\ Box 6730, SE-113 85 Stockholm, Sweden}

\vs{15mm}

{\bf Abstract }
\end{center}

\noindent
We describe three-dimensional Kerr-de Sitter space using similar methods as recently applied to the
BTZ black hole. A rigorous form of the classical connection between gravity in three dimensions and
two-dimensional conformal field theory is employed, where the fundamental degrees of freedom are
described in terms of two dependent $SL(2,\mathbb{C})$ currents. In contrast to the BTZ case,
however, quantization does not give the Bekenstein-Hawking entropy connected to the cosmological
horizon of Kerr-de Sitter space.

\np
\setcounter{page}{1}

\section{Introduction}
\label{sec:introduction}
The past few years have seen significant progress in the understanding of the fundamental degrees of
freedom in quantum gravity, most notably within string theory and regarding the calculation of black
hole entropy from statistical mechanics. The entropy of a black hole appears to be connected with
the presence of a horizon, but black holes are not the only geometries with horizons. It was shown~
\cite{GibHawk:cosmhor} early on that also cosmological horizons have corresponding entropies, given
by the same expression
\[S=\frac{A}{4G}\] (the Bekenstein-Hawking formula) as for black holes, where now $A$ is the (
generalized) area of the cosmological horizon and $G$ is Newton's constant. A prototype of
spacetimes with cosmological horizons is provided by de Sitter spacetime, and the spinning black
hole generalizations, Kerr-de Sitter (KdS).

Following Carlip~\cite{Carlip:95}, numerous papers have appeared calculating the entropy of the
three-dimensional BTZ~\cite{BTZ} black hole. As emphasized~\cite{Carlip:98}, these share some
serious shortcomings, and it looks difficult to arrive at the Bekenstein-Hawking formula from pure
gravity. In the light of these shortcomings it was proposed in~\cite{FjHw:01} how to reconcile pure
gravity with the Bekenstein-Hawking formula. A main theme in that work was the discovery of new
sectors of solutions which provided the necessary increase in the number of degrees of freedom. This
was in turn inspired by earlier results~\cite{FjHw:99} where a classical equivalence between Chern-
Simons gauge theory and a WZNW model was rigorously formulated, and also by the construction of
multi-black hole solutions~\cite{MaSu:00} in three-dimensional gravity with negative cosmological
constant.

Starting with Maldacena and Strominger~\cite{MaldStr:98} similar methods as those used on the BTZ
black hole have been employed for positive cosmological constant in three dimensions. There are no
black holes in this theory, but solutions in general have a cosmological horizon. The most recent of
these methods makes use of a conjectured correspondence between quantum gravity on a de Sitter
background and (Euclidean) conformal field theory (CFT) on the conformal boundary of de Sitter
spacetime~\cite{Str:dSCFT}. In this case the boundary is disconnected with one spacelike component
in the asymptotic past and one in the asymptotic future.
These methods share all the shortcomings of the original methods, in addition to new interpretative
difficulties due to the boundary being spacelike.

The main purpose of the present investigation was to investigate whether the results of~\cite{
FjHw:01} on the statistical mechanics of the BTZ black hole generalize to the case of positive
cosmological constant. We consider here a two-parameter family of solutions, the KdS$_{3}$
solutions first considered in~\cite{Park:KdS} where also a macroscopic derivation of the entropy is
presented.
Previous results depended crucially on the structure of unitary representations of $SL(2,\RR)$, as
being the gauge group in the Chern-Simons theory. In the present case the gauge group is
$SL(2,\CC)$, whose representations are drastically different from those of $SL(2,\RR)$. It is
therefore far from obvious that the entropy calculation for the BTZ black hole can be generalized to
KdS$_{3}$, and as we will show it indeed does not.

In the next section we establish the equivalence between three-dimensional gravity with positive
cosmological constant on a manifold with boundary, and a conformal field theory (CFT) closely
related to the $SL(2,\mathbb{C})$ WZNW model on the boundary. There is an important difference to
the dS/CFT correspondence, however, since the boundary is not the conformal boundary of spacetime.
Rather it is the boundary of a disc-like region excised around a charge which appears in the Chern-
Simons formulation of gravity. This boundary is in general timelike. The procedure is actually
independent of the location of the boundary, and the correspondence is merely a consequence of the
geometry having mass and/or spin.
Section~\ref{sec:holonomies} is devoted to a brief discussion of the topological observables of the
Chern-Simons theory, the Wilson loops.
We then establish the existence of multi-center solutions also for positive cosmological constant in
section~\ref{sec:multicenter}. The obstruction to generalizing the BTZ story is encountered in
section~\ref{sec:sectors} when we try to introduce new non-local sectors of solutions. Nevertheless
we proceed with the construction of the state space in section~\ref{sec:statespace}, and the vacuum
structure seem to conform with recent results in the dS/CFT correspondence We also show that the
statistical entropy does not coincide with the Bekenstein-Hawking entropy. Finally a brief
discussion is included, relating our results to some other methods.

\section{3d gravity with $\Lambda>0$ as CFT}
\label{sec:gravCSCFT}
From the seminal work of Witten~\cite{Witten:gravcs} we know that three-dimensional gravity can be
formulated as a Chern-Simons gauge theory. In the case of positive cosmological constant
$\Lambda=\frac{1}{l^{2}}$ the gauge group is $SO(3,1)\cong SL(2,\CC)$, and up to a boundary term the
Einstein-Hilbert action is equivalent to
\begin{eqnarray}
   I_{EH} & = & -\frac{k}{4\pi}\int_{\M}Tr\left[A\wedge dA+\frac{2}{3}A\wedge A\wedge A\right]+\frac
   {k}{4\pi}\int_{\M}Tr\left[\bar{A}\wedge d\bar{A}+\frac{2}{3}\bar{A}\wedge \bar{A}\wedge \bar{A}
   \right]\nonumber\\
   & = & \frac{\lambda}{2\pi}Im\left(\int_{\M}Tr\left[A\wedge dA+\frac{2}{3}A\wedge A\wedge A\right]
   \right)\nonumber\\
   \label{csaction}
   & = & I_{CS}[A]-I_{CS}[\bar{A}]
\end{eqnarray}
where
\begin{eqnarray}
   A & = & \left(\omega^{a}+\frac{i}{l}e^{a}\right)T_{a}\\
   \bar{A} & = & \left(\omega^{a}-\frac{i}{l}e^{a}\right)\bar{T}_{a}\\
   k & = & -\frac{il}{8G} : = i\la
\end{eqnarray}
The $\mathfrak{sl}(2,\mathbb{C})$ generators $T_{a}$ and $\bar{T}_{a}$ satisfy
\begin{eqnarray}
   \left[ T_{a}, T_{b}\right]  & = & f_{ab}^{\ \ c} T_{c}\\
   \left[\bar{T}_{a}, \bar{T}_{b}\right]  & = & f_{ab}^{\ \ c} \bar{T}_{c}\\
   \left[ T_{a}, \bar{T}_{b}\right] & = & 0
\end{eqnarray}
with $f_{ab}^{\ \ c} = \ep_{abd}\eta^{dc}$, $\eta^{ab}=diag(-1,1,1)$ and $\ep_{012}=+1$~\footnote{
Compared to the $SO(2,2)$ case in~\cite{FjHw:01} the normalisation of the generators is chosen
differently so that here $Tr(T_{a}T_{b}) = \eta_{ab}$. This explains the value of $k$ above.}. This
complex basis is related to a real basis of $\mathfrak{so}(3,1)$ via
\begin{eqnarray}
   M_{a} & = & T_{a}+\bar{T}_{a}\\
   P_{a} & = & \frac{i}{l}\left(T_{a}-\bar{T}_{a}\right)
\end{eqnarray}
where
\begin{eqnarray}
   \label{JPrel1}
   \left[ M_{a}, M_{b}\right]  & = & f_{ab}^{\ \ c} M_{c}\\
   \label{JPrel2}
   \left[ P_{a}, P_{b}\right]  & = & -\frac{1}{l^{2}}f_{ab}^{\ \ c} M_{c}\\
   \label{JPrel3}
   \left[ M_{a}, P_{b}\right]  & = & f_{ab}^{\ \ c} P_{c}.
\end{eqnarray}
To simplify the discussion of representations, let us introduce the following more common basis of
$\mathfrak{so}(3,1)$
\[L_{MN}=-L_{NM}, \quad N,M=0,\ldots,3,\quad \eta_{MN} =diag(-1,1,1,1)\] in which the commutation
relations read
\begin{equation}
   \label{lorentzcomm}
   \left[L_{KL},L_{MN}\right] = \eta_{KM}L_{LN} + \eta_{LN}L_{KM} - \eta_{KN}L_{LM} -
   \eta_{LM}L_{KN}.
\end{equation}
This is the natural basis when utilizing $SO(3,1)$ as the Lorentz group in four-dimensional
spacetime. Two quadratic Casimirs are easily constructed in this basis
\begin{eqnarray}
   C_{1} & = & \frac{1}{2}\eta^{KM}\eta^{LN}L_{KL}L_{MN}\\
   & = & -L_{12}^{2} - L_{23}^{2}-L_{31}^{2}+L_{01}^{2}+L_{02}^{2}+L_{03}^{2}\\
   C_{2} & = & \frac{1}{4}\varepsilon^{KLMN}L_{KL}L_{MN}\\
   & = & L_{01}L_{23}-L_{12}L_{30}+L_{20}L_{13}
\end{eqnarray}
and the relation to the $M,P$ basis is given as
\[M_{0}=L_{12},\ M_{1}=L_{20},\ M_{2}=L_{01}\]\[P_{0}=l^{-1}L_{30},\ P_{1}=l^{-1}L_{13},\ P_{2}=l^{-
1}L_{23}.\]
Thus
\begin{eqnarray}
   C_{1} & = & \eta^{ab}M_{a}M_{b}-l^{2}\eta^{ab}P_{a}P_{b}\\
   C_{2} & = & l\eta^{ab}M_{a}P_{b}
\end{eqnarray}
and
\begin{eqnarray}
   \eta^{ab}T_{a}T_{b} & = & \frac{1}{4}(C_{1}+2iC_{2})\\
   \eta^{ab}\bar{T}_{a}\bar{T}_{b} & = & \frac{1}{4}(C_{1}-2iC_{2}).
\end{eqnarray}
Any unitary representation $D$ of $\mathfrak{so}(3,1)$ can be decomposed into a direct sum of
unitary representations, $D_{n}$, of the compact $\mathfrak{su}(2)$ algebra generated by $L_{ij}$,
$i, j=1,2,3$. The label $n$ is the usual spin. A unitary irreducible representation of
$\mathfrak{so}(3,1)$ carry a lowest spin $k_{0}$, but since unitary irreducible representations are
infinite dimensional there is no upper bound, i.e. \[D=\bigoplus_{n=k_{0}\atop
step=1}^{\infty}D_{n}.\] Note that if $k_{0}$ is an integer the full $\mathfrak{so}(3,1)$
representation will only contain integer $\mathfrak{su}(2)$ spins, and if $k_{0}$ is a half-integer
it will only involve half-integral spins.
The unitary irreducible representations $D_{c_{1}}^{c_{2}}$ come in two series labelled by the
values of the Casimirs~\cite{Bargmann:46, Knapp:repth}
\begin{enumerate}
   \item The complementary series $D^{0}_{c_{1}}$ where $0<c_{1}\leq 1$ and $c_{2}=0$
   \item The principal series $D^{c_{2}}_{c_{1}}$ where $c_{2}\in\RR$ and
   $c_{1}=k_{0}^{2}-(c_{2}/k_{0})^{2}-1$ for $k_{0}\in\frac{1}{2}\mathbb{N}$.
\end{enumerate}
The identity representation in fact corresponds to the upper limit of the complementary series, i.e.
$D_{1}^{0}$.
We can also classify the unitary irreducible repreentations by the lowest spin $k_{0}$ together with
a parameter $c$~\cite{Naimark:linreps} related to the Casimirs as
\begin{eqnarray}
   c_{1} & = & k_{0}^{2}+c^{2}-1\\
   c_{2} & = & -ik_{0}c
\end{eqnarray}
The two series correspond to
\begin{enumerate}
   \item $D^{0}_{c_{1}}$: $k_{0}=0$ and $0<c<1$
   \item $D^{c_{2}}_{k_{0}^{2}-(c_{2}/k_{0})^{2}-1}$: $c=i\rho$, $\rho\in\RR$ and
   $k_{0}\in\frac{1}{2}\mathbb{N}$.
\end{enumerate}

Diagonalize the operator $M_{0}$ and denote its eigenvalues in $D_{j}$ by $m_{j}$ where
$m_{j}\in\{-j,-j+1,\ldots,j-1,j\}$. There is no compact generator which commutes with $M_{0}$ so the
other Cartan generator must be a non-compact  generator, which then has continuous eigenvalues. The
eigenvalue of the operator $P_{0}$, which commutes with $M_{0}$, is given by \[\frac{i}{l}(m+k_{0}+c
+1+\si)\] where $\si\in\RR$ is arbitrary.

Since the action (\ref{csaction}) consists of two commuting parts, the canonical analysis on a
manifold ${\M} =\RR\times\Si$ such that $\p\Si\neq\emptyset$ can be directly transfered from the
$SO(2,2)$ case, and for details we refer to~\cite{FjHw:99} and~\cite{FjHw:01}. The boundary destroys
gauge invariance, and to restore this we introduce two sets of $\sltwo$ currents on the boundary,
parametrized by the coordinate $\phi$, with the equal time Poisson relations
\begin{eqnarray}
   \label{Jalg1}
   \{J_{a}(\phi),J_{b}(\phi')\} & = & -f_{ab}^{\ \ c}J_{c}(\phi)\del(\phi-\phi') - \frac{k}{4\pi}
   \eta_{ab}\p_{\phi}\del(\phi-\phi')\\
   \label{Jalg2}
   \{\bar{J}_{a}(\phi),\bar{J}_{b}(\phi')\} & = & -f_{ab}^{\ \ c}\bar{J}_{c}(\phi)\del(\phi-\phi') +
   \frac{k}{4\pi}\eta_{ab}\p_{\phi}\del(\phi-\phi').
\end{eqnarray}
Including these boundary degrees of freedom we can write down first class constraints of a form
similar to those in~\cite{FjHw:01}
\begin{equation}
\label{constraints}
\p_{r}A_{\phi}^{\ a}-\p_{\phi}A_{r}^{\ a} + f^{a}_{bc}A_{r}^bA_{\phi}^c-(A_{\phi}^{\
a}+\frac{4\pi}{k}J^{a})\delta(x\in\p\Sigma)=0
\end{equation}
where $r$ is chosen such that $(r,\phi)$ parametrize $\Si$.
The boundary delta function $\delta(x\in\p\Sigma)$ is defined by
$\int_{\Sigma}d^2xf(x)\delta(x\in\p\Sigma)=\int_{\p\Sigma}[f(x)]_{\p\Sigma}$.
We refer to ~\cite{FjHw:01} for details of the canonical analysis.
On the constraint surface, and considering only smooth field configurations near and on the
boundary, these constraints imply
\begin{eqnarray}
   J^{a} & = & -\frac{k}{4\pi}A_{\phi}^{\ a}|_{\p\Si}\\
   \bar{J}^{a} & = & \frac{k}{4\pi}\bar{A}_{\phi}^{\ a}|_{\p\Si}.
\end{eqnarray}
Since $(A^{a})^{*}=\bar{A}^{a}$, the currents satisfy \begin{equation}\label{JJbar} (J^{a})^{*}=\bar
{J}^{a}.\end{equation}

From~\cite{FjHw:99} and~\cite{FjHw:01} we know that we can fix the gauge and obtain the following
gauge fixed Hamiltonian
\begin{equation}
   \label{Ham}
   H = \frac{1}{2\al kl}\int_{\p\Si}J^{a}J_{a} +
   \frac{1}{2\bar{\al}kl}\int_{\p\Si}\bar{J}^{a}\bar{J}_{a}
\end{equation}
where $\al$ and $\bar{\al}$ are arbitrary constants with the constraint $\al\neq-\bar{\al}$~
\footnote{As explained in~\cite{FjHw:01}, the case $\al=-\bar{\al}$ yields a doubly chiral
Hamiltonian. A symmetry argument shows that this is not the correct choice for gravity, and the
argument works equally well in the present setting.}.
We fix these parameters such that
\begin{eqnarray}
   \label{wzwH}
   lH & = & -\frac{2\pi}{\lambda}\int_{\p\Si}J^{a}J_{a} -
   \frac{2\pi}{\lambda}\int_{\p\Si}\bar{J}^{a}\bar{J}_{a}\\
   & = & L_{0}+\bar{L}_{0}.
\end{eqnarray}
The Fourier modes of the currents are defined as \begin{eqnarray} \label{Fmodes} J^{a} & = &
\frac{1}{2\pi}\sum_{m\in\mathbb{Z}}J^{a}_{m}e^{-im\phi}\\ \label{barFmodes} \bar{J}^{a} & = & \frac{
1}{2\pi}\sum_{m\in\mathbb{Z}}\bar{J}^{a}_{m}e^{im\phi}.\end{eqnarray}

Inserting the Fourier decomposition in (\ref{wzwH}) we get
\begin{eqnarray}
   L_{0} & = & \frac{1}{4\lambda}\left(C_{1}+2iC_{2}\right)+ \mbox{non-zero mode part}\\
   \label{LCas}
   \bar{L}_{0}  & = & \frac{1}{4\lambda}\left(C_{1}-2iC_{2}\right)+ \mbox{non-zero mode part}
\end{eqnarray}
In the case of negative cosmological constant the two parts of the Hamiltonian combine to the full
$SL(2,\RR)$ WZNW Hamiltonian. Here, however, we do not obtain the full $SL(2,\CC)$ WZNW model since
each current do not independently generate the $\mathfrak{sl}(2,\CC)$ algebra. Instead the classical
degrees of freedom correspond to those of the so called $H_{3}^{+}$ WZNW model~\cite{Gawedzki:91,
Teschner:97_1, Teschner:97_2, GivKutSeib:98, Teschner:99}, conjectured to describe string
propagation on the Euclidean version of AdS$_3$.

There exists a two-parameter class of three-geometries called Kerr-de Sitter (KdS$_{3}$)
solutions~\cite{Park:KdS} with metrics
\begin{equation}
   \label{KdS}
   ds^{2} = -\left(GM-\frac{r^{2}}{l^{2}}+\frac{G^{2}\J^{2}}{4r^{2}}\right)dt^{2} + \left(GM-\frac{r
   ^{2}}{l^{2}}+\frac{G^{2}\J^{2}}{4r^{2}}\right)^{-1}dr^{2} + r^{2}\left(d\phi - \frac{G\J}{2r^{2}}
   dt\right)^{2}.
\end{equation}
The case $M=1/G$, $\J=0$ corresponds to pure de Sitter which has a cosmological horizon at
$r=r_{+}=l$. For generic values of the parameters there is a single cosmological horizon, thus one
should not speak of black holes. The horizon dissapears, however, for $M=\J =0$. By inspection, it
is also clear that there is, in general, a conical singularity at $r=0$.
The parameters $M$ and $\J$ can be expressed in terms of the parameters $r_{+}$ and $r_{-}$ as \[M=
\frac{r_{+}^{2}-r_{-}^{2}}{l^{2}G} \quad \mbox{and} \quad\J=\frac{2r_{+}r_{-}}{lG}\] analogous to
the expressions for the BTZ case.

In the region $r>r_+$ we can perform the following coordinate transformation
\begin{equation}
r^2=r_+^2\cosh^2{\tau}+r_-^2\sinh^2{\tau}
\end{equation}
The metric then transforms to
\begin{equation}
\label{kerr-de metric}
ds^2 = -l^2 d\tau^{2} + \cosh^2{\tau}\left(\frac{r_-}{l}dt-r_+d\phi\right)^2
       +\sinh^2{\tau}\left(\frac{r_+}{l}dt+r_-d\phi\right)^2 \
\end{equation}
The dreibein and spin connections can now be chosen to be
\begin{align}
   e_{0} & =  l d\tau & \omega_{0} & = 0 \\
   e_{1} & = -\sinh{\tau}(\frac{r_+}{l}dt+r_-d\phi) & \omega_{1} & = \frac1{l}
   \sinh{\tau}(\frac{r_-}{l}dt-r_+d\phi)  \\
   e_{2} & =  -\cosh{\tau}(\frac{r_-}{l}dt-r_+d\phi)  & \omega_{2} & = -\frac1{l} \cosh{\tau}(\frac{
   r_+}{l}dt+r_-d\phi).
\end{align}
By the relation $A_\al=\omega_\al+\frac{i}{l}e_\al$  the $\phi$-component of the vector field
corresponding to this metric is
\begin{equation}
   A_{\phi} = -\frac1{l}\left(\sinh\tau\, [r_{+}+ir_{-}]T_{1} + \cosh\tau\, [r_{-}-ir_{+}]T_{2}
   \right)
\end{equation}
which yields the currents
\begin{eqnarray}
   J^{0} & = & 0\\
   J^{1} & = & \frac{i\lambda}{4\pi l}\sinh\tau [r_{+}+ir_{-}]\\
   J^{2} & = & \frac{i\lambda}{4\pi l}\cosh\tau [r_{-}-ir_{+}].
  \end{eqnarray}
The $\bar{J}$ sector is obtained by complex conjugation, and the Virasoro zero-modes become
\begin{eqnarray}
   \label{LzMJ1}
   L_{0} & = & -\frac{\lambda}{4 l^2}\left[ r_{+}^{2}-r_{-}^{2}+2ir_{+}r_{-}\right]\\
   \label{MJ1}
   & = & -\frac{\lambda G}{4}\left[ M-i\J/l \right]\\
   \label{LzMJ2}
   \bar{L}_{0} & = & -\frac{\lambda}{4 l^2}\left[ r_{+}^{2}-r_{-}^{2}-2ir_{+}r_{-}\right]\\
   \label{MJ2}
   & = & -\frac{\lambda G}{4}\left[ M+i\J/l\right].
 \end{eqnarray}
Comparison with (\ref{LCas}) yields
\begin{eqnarray}
   M & = & -\frac{1}{G\lambda^{2}}C_{1}\\
   & = & -\frac{1}{G\lambda^{2}}(k_{0}^{2}+c^{2}-1)\\
   \J & = & \frac{2\, l}{G\lambda^{2}}C_{2}\\
   & = & -\frac{i2\, l}{G\lambda^{2}}k_{0}c.
\end{eqnarray}
We see in particular that all geometries with positive mass correspond to principal representations.
The relevance of the negative mass geometries is not clear to us.

\section{Holonomies}
\label{sec:holonomies}
The topological nature of Chern-Simons theory implies that the fields are locally pure gauge
\(A = U^{-1}dU\), for \( U \) an element of \(SL(2,\mathbb{C})\),
and any non-trivial observable has to be associated with the boundaries
of spacetime or be topological. The simplest topological
observables are holonomies (or Wilson loops) measuring the effect of
parallel transport along a closed loop in spacetime. For flat
connections the result can only be non-zero if the loop \( C \)  is
non-contractible. Then the Wilson loop is
\begin{equation}
     W(C)={\mathcal{P}}\exp\left(\oint_{C}A\right)
\end{equation}
where \( {\mathcal{P}} \) denotes path ordering of the exponential. The Wilson loop by itself is not
invariant under gauge transformations but the trace of it is.

For the KdS$_{3}$ solutions we obtain~\footnote{The normalization factor of $4$ is due to the choice
of representation.}
\begin{eqnarray}
    {\mathrm{Tr}}W(C) & = & 4 \left[\cosh \left(\pi\frac{ r_-}{l}\right)\cos\left(\pi \frac{r_+}{l}
    \right)-
    i\sinh\left(\pi \frac{r_-}{l}\right)\sin\left(\pi \frac{r_+}{l}\right)\right]\\
    \label{KdSWl}
    & = & 4\cosh\left(\pi\frac{r_{-}-ir_{+}}{l}\right)\\
    {\mathrm{Tr}}\bar{W}(C) & = & 4\left[\cosh\left(\pi \frac{r_-}{l}\right)\cos\left(\pi
    \frac{r_+}{l}\right)
    +i\sinh\left(\pi \frac{r_-}{l}\right)\sin\left(\pi \frac{r_+}{l}\right)\right]\\
    & = & 4\cosh\left(\pi\frac{r_{-}+ir_{+}}{l}\right)
    \label{eq:W1}
\end{eqnarray}
which  relate the holonomies to the mass and spin of the geometry. So we see that generically the
Wilson loop will take {\em complex} values, and genuine observables are related to real combinations
of traces of the Wilson loops. This in turn implies that every possible eigenvalue of the Wilson
loop corresponds to a certain KdS$_{3}$ solution which is not the case for the BTZ solution.
It may be interesting also to notice the value of the holonomy for pure de Sitter. As mentioned
earlier, this is obtained by setting $r_+=l$ and $r_-=0$. Then we see that the trace of the holonomy
takes the value ${\mathrm{Tr}}W(C)=-4$.
A peculiar feature is the periodicity in $r_{+}$.

In analogy to the case of negative curvature there also exist multicenter solutions when the
 curvature is positive which will be established in the next session.
For these solutions the Wilson loop will just add up all separate charges ($r_+$ respective $r_-$
for each source) that are enclosed~\cite{MaSu:00}. If we enclose  KdS$_{3}$-like solutions the
Wilson loops are
\begin{eqnarray}
    {\mathrm{Tr}}W(C) & = & 4\cosh\left(\pi\frac{r_{C-}-ir_{C+}}{l}\right)\\
    {\mathrm{Tr}}\bar{W}(C) & = & 4\cosh\left(\pi\frac{r_{C-}+ir_{C+}}{l}\right),
    \label{eq:W2}
\end{eqnarray}
where $r_{C+}$ ($r_{C-}$) denotes the sum of all charges $r_+$ ($r_-$) enclosed by $C$. Here we see
an important difference between the KdS$_{3}$ and the BTZ black hole case. If several spinless (i.e.
$r_{-}=0$) KdS$_{3}$ sources are encircled such that the sum of all $r_+$ adds up to $l(1+2n)$ for
some integer $n$, then by the periodicity the total Wilson loop at infinity will be that of pure de
Sitter.

\section{Multi-center solutions}
\label{sec:multicenter}
In \cite{MaSu:00} it was shown that in the Chern-Simons formulation there exist multi-center
generalizations of the BTZ-solution, and in \cite{FjHw:01} the solutions were further generalized to
line sources. It is
easy to generalize the multi-center solutions to positive cosmological constant. Here we will write
down the solution for the case of having point sources and we will also ensure that it behaves
asympotically as KdS$_{3}$. We do, however, leave an analysis of the physical relevance, as
performed in \cite{MaSu:00}, aside.
For point source solutions the equations of motion $dA + A \wedge A=0$ are satisfied by the vector
potential outside the sources.
We previously used these solutions as inspiration for the introduction of new sectors of solutions,
and in that case we believe they provide an important contribution to the density of states.

It should be pointed out that even though inside the horizon the radial coordinate $\tau$ is spatial
also for KdS$_{3}$, the asymptotic behavour is quite different. Outside the horizon $\tau$ becomes
the time coordinate, and the boundary at infinity is spacelike.
The vector field for such a solution outside the horizon can be written
\begin{equation}
\label{ansatz1}
A=-(f+Qdt)\sinh(h)T_1 + i(f+Qdt)\cosh(h)T_2-dh T_0
\end{equation}
where \( h \) is a scalar function generalizing the  coordinate
$\tau$ and \( f \) is a  one-form inside the horizon which is closed except at
isolated sources
\begin{equation}
    df = 2 \pi\sum_{i=1}^{N} q_i \delta^2(\vec{x}-\vec{x}_i)\; dx\wedge
    dy~.\label{eq:f}
\end{equation}
The charges $q_{i}=r_{i+}-ir_{i-}$ determine the strength of the sources (the masses
and spins of KdS$_{3}$ solutions). By integrating (\ref{eq:f}) over a large
disk \( D \) enclosing all sources we obtain
\begin {equation}
\oint_{\partial D} f =\int_{D} df = 2 \pi \sum_{i=1}^{N} q_i  = 2\pi Q~.
\end{equation}
If appropriate boundary conditions on \( f \) are assumed, \( f \to Q
d\phi \) as \( r \to \infty \). The second gauge field \(\bar{A}\) is
 just the complex conjugate of $A$.

The metric corresponding to the Chern-Simon fields is,
\begin{equation}
ds^2=\cosh^2(h(x,y))\left\{r_- dt - \Re(f(x,y))\right\}^2
    +\sinh^2(h(x,y))\left\{ r_+ dt + \Im(f(x,y)) \right\}^2
        -dh(x,y)^2~,\label{SimpMultiBTZMet}
\end{equation}
where
\begin{equation}
    r_{+}=\Re(Q) ~,\qquad r_{-}=\Im(Q)
\end{equation}
The metric is easily compared with (\ref{kerr-de metric}). The one-form \( f /Q \) generalizes the
angular one-form \( d\phi \).
Asymptotically it behaves like KdS$_{3}$ as long as $h(x,y)\rightarrow \tau$ at infinity.

It may appear strange to discuss spatial sections with boundary when we intend to describe KdS$_{3}$
geometries which have no spatial boundary. The reason is that we insist on describing the whole
$\Si$ with one set of coordinates. As described in~\cite{FjHw:01} all (but a zero-measure set of)
solutions are then described by gauge connections which are singular along some world-line in the
Chern-Simons theory. This is made obvious by the discussion above regarding the construction of
multi-center solutions, where the position of the charges correspond to such singularities.
A singularity forces us to remove a (small) disc around the charge in each spatial section,
presenting us with a boundary. In a multi-center solution we should remove a disc around each charge
which results in several disconnected bondaries, $\p\Si_{i}$, each supporting a current $J^{a}_{i}$.
We thus stress that the presence of boundaries in $\mathcal{M}$ is a natural feature.

Another important feature is that nothing in the formalism depends on the location of the boundary,
and it is strictly not correct to say that the CFT lives "on the boundary", whether that is at
infinity or any other place. Thus although we are forced to introduce boundaries, {\em the locations
of the boundaries are irrelevant}.

One might think that an infinite number of currents are needed to describe all possible
configurations, but this is interestingly enough not the case. It is possible~\cite{FjHw:99} to
choose a gauge in the full theory such that
\[ J^{a} = -\frac{k}{4\pi}A_{\phi}^{\ a} \] {\em everywhere}, and not only on the boundary. For a
solution with several disconnected boundaries this constraint relates all currents to each other,
and only one is really needed. Note, however, that it may well be impossible to apply this
constraint globally, and in the previous work~\cite{FjHw:01} this provided inspiration to include
also sectors with non-local boundary conditions.

\section{Generating solutions by singular gauge transformations}
\label{sec:sectors}
We will now investigate what solutions can be generated by singular gauge transformations. In~\cite{
FjHw:01} this technique was used to find new sectors of solutions. We consider transformations of
$B=A+\bar{A}$. Consider the one-parameter family of transformations
\[ B\Rightarrow g^{-1}(B+d)g,\quad g=e^{-\phi(sT_{0}+\bar{s}\bar{T}_{0})}.\]
We obtain
\[\tilde{A}_{\al}^{\ 0} = A_{\al}^{\ 0}-s\del_{\al,2},\quad \tilde{A}^{\ \pm}_{\al} = e^{\mp is\phi}
A_{\al}^{\ \pm}\]
\[\tilde{\bar{A}}_{\al}^{\ 0} =\bar{A}_{\al}^{\ 0} - \bar{s}\del_{\al, 2},\quad
\tilde{\bar{A}}_{\al}^{\ \pm} = e^{\mp i s\phi}\bar{A}_{\al}^{\ \pm}\]
where $A^{\pm}=iA^{1}\pm A^{2}$ and vice versa for $\bar{A}$.
Demanding that the transformed fields are still related by complex conjugation gives the constraint
$$ s^*=\bar{s}$$ with two special solutions $$s\in \RR,\  s=\bar{s}$$ and $$is\in\RR,\ s=-\bar{s}.$$
In the first case the group element $g$ used in the transformation becomes $g=e^{-s\phi M_{0}}$
which is periodic with period $2\pi$, and therefore a regular gauge transformation, for
$s\in\mathbb{Z}$.
The second case implies $g = e^{is\phi P_{0}}$ which is only a regular gaugetransformation for $s=0$
since $P_{0}^{\,\dagger}=P_{0}$.
Extending this to a complete "gauge" transformation, i.e. transforming also the boundary currents,
gives
\begin{eqnarray}
   \tilde{J}^{0} & = & J^{0}+\frac{sk}{4\pi},\quad \tilde{J}^{\pm} = e^{\mp is\phi}J^{\pm}\nonumber
   \\
   \label{Jtrf}
   \tilde{\bar{J}}^{0} & = & \bar{J}^{0}-\frac{\bar{s}k}{4\pi},\quad \tilde{\bar{J}}^{\pm} = e^{\mp
   i\bar{s}\phi}\bar{J}^{\pm}
\end{eqnarray}
or in terms of Fourier modes
\begin{eqnarray}
   \tilde{J}^{0}_{n} = J^{0}_{n}+\frac{sk}{2}\del_{n},&& \tilde{J}^{\pm}_{n} = J^{\pm}_{n\mp s}
   \nonumber\\
   \label{mtrf}
   \tilde{\bar{J}}^{0}_{n} = \bar{J}^{0}_{n}-\frac{\bar{s}k}{2}\del_{n},&& \tilde{\bar{J}}^{\pm}_{n}
   = \bar{J}^{\pm}_{n\mp \bar{s}}
\end{eqnarray}
which we recognize as the spectral flow of $\widehat{\mathfrak{sl}}_{2}$.
To see how the $\mathfrak{so}(3,1)$ representations are affected we move to the $M,\,P$-basis. Since
$M_{0}^{0}= J^{0}_{0}+\bar{J}^{0}_{0}$ is a compact generator it has discrete eigenvalues while
$P_{0}^{0}=i(J^{0}_{0}-\bar{J}^{0}_{0})$, being non-compact, has continuous spectrum. Therefore,
following~\cite{FjHw:01}, we should consider $M_{0}^{0}$ which transforms as $$M_0^0\rightarrow
\widetilde{M}_0^0=M_0^0+\frac{k}{2}(s-\bar{s}).$$ The first case, $s$ real, implies that $M_0^0$
does not transform at all, and consequently there is no restriction on $s$ from representation
theory. In the second case, $s$ imaginary, we have $$M_0^0\rightarrow \widetilde{M}_0^0=M_0^0+ks.$$
Performing this transformation on a unitary irreducible representation of $\mathfrak{so}(3,1)$ has
the effect of shifting the $M_0$ eigenvalue, and to stay within the class of unitary representations
we must demand $ks\in\frac{1}{2}\mathbb{Z}$. Unlike the BTZ case, however, there are no values of
$s$ which give regular gauge transformations, and consequently $k$ is not quantized and there is no
bound on the number of new sectors.
There are also further differences regarding the solution generating transformations above.
The values of the KdS$_3$ Wilson loops (\ref{KdSWl}) changes according to
\begin{equation}\label{Wtransf}{\mathrm Tr} W[C]\longrightarrow 4\cosh \left(\pi\frac{r_{-}-ir_{+}}{
l}+i\pi s\right). \end{equation}
These transformations do not generate new eigenvalues of the Wilson loops since any value of $s$ can
be incorporated in a redefinition of $r_{+}$ and $r_{-}$ while remaining in the class of KdS$_3$
solutions. This is in contrast to the AdS$_3$ case where the analogous transformation generated
genuinely new Wilson loops.

We have discovered a significant difference between the space of solutions for positive and negative
cosmological constant in three dimensions. An important consequence is that the microscopic
mechanisms responsible for the entropy are not the same for positive $\La$ as for negative $\La$. It
should be pointed out that the result obtained in this sector is independent of the spectral flow
direction in the $\widehat{\mathfrak{sl}}_2$ algebras.

\section{Quantum state space}
\label{sec:statespace}
Since the left- and right-moving currents are now related by complex conjugation, it is perhaps not
obvious how to construct the state space from the Fourier modes. By moving to the real $M$, $P$
basis, it is obvious that there are $6$ real field degrees of freedom, and we should thus let
$J^{a}$ and $\bar{J}^{a}$ act independently. The relation (\ref{JJbar}), which in the quantized
version reads \[(J^{a}_{n})^{\dagger} = \bar{J}^{a}_{n},\] has instead implications for the inner
product of the state space as will now be discussed.
We have in principle two possible choices of vacua, either
\[ J^{a}_{n}|0\rangle = \bar{J}^{a}_{-n}|0\rangle = 0 \quad \forall n>0 \] or \[J^{a}_{n}|0\rangle =
\bar{J}^{a}_{n}|0\rangle = 0\quad \forall n>0.\]
The commutators corresponding to (\ref{Jalg1}) and (\ref{Jalg2}) read in modes
\begin{eqnarray}
   \label{qJalg1}
   [J^{a}_{m},J^{b}_{n}] & = & -if^{ab}_{\ \ c}J^{c}_{m+n} -\frac{k}{2}m\del_{m+n,0}\eta^{ab} \\
   \label{qJalg2}
   [\bar{J}^{a}_{m},\bar{J}^{b}_{n}] & = & -if^{ab}_{\ \ c}\bar{J}^{c}_{m+n} -\frac{k}{2}m\del_{m+
   n,0}\eta^{ab}.
\end{eqnarray}
In the first choice of state space we define a sesquilinear form by
\[ (|\psi\rangle,|\phi\rangle ) = \langle\psi|\phi\rangle \] where
$\langle\psi| = (|\psi\rangle )^{\dagger}$. This is the first step in constructing an inner product
on the state space, and it is instructive to study a few simple examples. By defining
$|1,a\rangle=J^{a}_{-1}|0\rangle$ and $|\bar{1},a\rangle=\bar{J}^{a}_{1}|0\rangle$ we get using the
hermiticity properties of the modes together with (\ref{qJalg1}) and (\ref{qJalg2})
\[
\begin{array}{ll}
   (|1,a\rangle,|1,b\rangle) = 0 & (|1,a\rangle,|\bar{1},b\rangle)= \frac{k}{2}\eta^{ab} \\
   (|\bar{1},a\rangle,|1,b\rangle) = -\frac{k}{2}\eta^{ab} & (|\bar{1},a\rangle,|\bar{1},b\rangle) =
   0.
\end{array}
\]
We see that the form is not diagonal in this basis, but it is still obvious that (even if we get rid
of the factor of $i$ present in $k$) the indefinite metric $\eta^{ab}$ results in an indefinite
sesquilinear form. We can thus not create a unitary theory in this state space.

If we try to define the form $( , )$ in the same way for the latter choice of vacuum, however, we
run into trouble. Since now
$(|\bar{1},a\rangle,|1,b\rangle) = \langle 0|J^{a}_{1}J^{b}_{-1}|0\rangle =
-\frac{k}{2}\eta^{ab}\langle 0|0\rangle$ and $|\bar{1},a\rangle=\bar{J}^{a}_{1}|0\rangle=0$, we
necessarily get $\langle 0|0\rangle =0$. Hence we need to define another state
$\widetilde{|0\rangle}$ by \[J^{a}_{-n}\widetilde{|0\rangle} = \bar{J}^{a}_{-n}\widetilde{|0\rangle}
= 0 \quad \forall n>0\] and then it is consistent to impose the normalization
$\widetilde{\langle 0|}0\rangle = 1$ and define the sesquilinear form
\[ (\widetilde{|\psi\rangle}, |\phi\rangle ) = \widetilde{\langle\psi|}\phi\rangle.\]
Since e.g.
$(\widetilde{|1,a\rangle},|2,b\rangle) = \frac{k}{2}\eta^{ab}\widetilde{\langle 0|}0\rangle$,
however, it is impossible to build a unitary state space also with this choice of vacuum.
It is interesting to note that the definition of the vacuum state in QFT on a de Sitter background
has recently been discussed~\cite{Witten:01, BouMalStr:01}, and seems to fit the second choice
above.

In certain applications, such as strings described by WZNW models with compact groups, unitarity
picks the latter vacuum. This is because the representations of the left- and right-moving affine
algebras then have the same value of the level $k$. With the former choice of vacuum, the levels in
the two chiral sectors differs by a sign. Since in that case the left- and right-moving currents are
not related by complex conjugation, the subtlety with the state $\widetilde{|0\rangle}$ does not
appear, however.
In the present gravitational context, the question of which vacuum to choose is not as obvious.
Consider the action of $H$ on a state corresponding to either of the two choices of vacuum.
Disregarding the zero-modes, we see that the commutation relations (\ref{qJalg1}) and (\ref{qJalg2})
implies the Hamiltonian acting on the former space yields $N-\bar{N}$ while acting on the latter
space it gives $N+\bar{N}$, where $N$ and $\bar{N}$ are the number operators in the respective
sectors. This, we believe, shows that the latter state space is more natural.

Since we could not establish the presence of different sectors this is the full story and the
calculation of the semiclassical entropy is a simple exercise.
The Hamiltonian is written in eq.~(\ref{wzwH}), and the state space is constructed by acting with
negative frequency modes on ground states $|R\rangle |\bar{R}\rangle$, transforming in some unitary
representation of $SL(2,\mathbb{C})$.
Recall that describing the theory only in terms of the currents involves a complete gauge fixing,
and there are therefore no constraints left to kill states in the state space constructed by
negative frequency modes of the six current degrees of freedom. A consequence of this and the fact
that the state space has the same number of states as that of three free bosons is that the
number of different states of a given chirality at "mode number" $N$ is simply the $3$rd partition
of $N$, and the asymptotic behaviour of this function is easily determined by e.g. a saddle point
method to be~\cite{Brigham} $$p^{(3)}(N)\sim e^{2\pi\sqrt{N/2}}.$$
Note that no assumptions on unitarity or modular invariance have been made.
For fixed values of the zero modes of $L_0$ and $\bar{L}_0$ the asymptotic number of states with (
large) eigenvalues
$\Delta$ and $\bar{\Delta}$ of $L_0$ and $\bar{L}_0$ is
\begin{equation}
   \label{density}
   \varrho(\Delta,\bar{\Delta})\sim e^{2\pi\sqrt{\Delta\over 2}}e^{2\pi\sqrt{\bar{\Delta}
   \over 2}}.
\end{equation}
We now assume that we can make the identification (\ref{MJ1}), (\ref{MJ2})
\begin{eqnarray*}
   \Delta & = & \frac{1}{16|k|}\left(\frac{r_+ + ir_-}{2G}\right)^{2}\\
   \bar{\Delta} & = & \frac{1}{16|k|}\left(\frac{r_+ - ir_-}{2G}\right)^{2}
\end{eqnarray*}
for an {\em arbitrary} state in the state space.
Inserted in (\ref{density}) this gives
\begin{equation}
   \label{entropy}
   S = \frac{1}{\sqrt{32|k|}} \frac{A}{4G}
\end{equation}
which in the semiclassical limit $|k|\rightarrow \infty$ is very far from the Bekenstein-Hawking
expression.
The partition function also contains an integration over the zero-modes, but it is straightforward
to check that this gives corrections to (\ref{entropy}) which are at most logarithmic in $\Delta$
and $\bar{\Delta}$.

\section{Discussion}
\label{sec:discussion}
Let us first give a short summary of the results. By using the Chern-Simons formulation of
three-dimensional gravity we showed that on a manifold $\mathcal{M}\cong \RR\times\Sigma$ where
$\p\Si\neq\emptyset$, three-dimensional gravity with positive cosmological constant is canonically
equivalent to two $SL(2,\CC)$ currents of opposite chirality and related by complex conjugation. We
furthermore determined the relation between the parameters of the vacuum (KdS$_3$) solutions, the
mass and spin, and the Casimirs of unitary irreducible representations of $SL(2,\CC)$. All positive
mass geometries were found to correspond to the principal unitary series. The eigenvalues of the
holonomies of these solutions were then calculated in terms of the mass and spin. Existence of "
multi-center" KdS$_3$ solutions were established in analogy with the multi-center black hole
solutions for $\La<0$. In the latter case the existence of multi-center solutions
served as motivation for introducing new sectors of solutions. We showed in
section~\ref{sec:sectors}, however, that similar new sectors do not appear for the present case
$\La>0$.
We then proceeded by showing that with the choice of vacuum implied by the gravitational
interpretation we are forced to introduce a new state, conjugate to the vacuum.
The state space will in agreement with the $\La<0$ case necessarily contain states of negative norm.
Finally we determined the asymptotic density of states to show that the space of solutions is not
degenerate enough to saturate the Bekenstein-Hawking entropy corresponding to the cosmological
horizon.

Although, as we have shown, much of the formalism is readily transferred between the
$\La<0$ and $\La>0$ cases, the spaces of solutions are quite different in structure. For
negative cosmological constant we found a mapping from the space of classical solutions to the state
space of the WZNW model such that the microcanonical density of states yields the Bekenstein-Hawking
entropy. Here we attempted the same construction for positive cosmological constant, and we found
that such a mapping does not give the Bekenstein-Hawking entropy corresponding to the cosmological
horizon. Although multi-center solutions exist in both cases, the relevance of these do not seem to
be the same.

Let us finish with a brief discussion of the relation to other results. In~\cite{MaldStr:98} a
method parallell to that of Carlip~\cite{Carlip:95} for the BTZ black hole was employed to calculate
the dS$_{3}$ entropy. The Hamiltonian in~\cite{MaldStr:98} reads using our conventions
\begin{equation}
   \label{MalStrHam}
   lH = \frac{1}{2k-1}\sum_{n\in\mathbb{Z}}:J^{a}_{n}J^{b}_{-n}:\eta_{ab} - \frac{1}{2k+1}\sum_{n\in
\mathbb{Z}}:\bar{J}^{a}_{n}\bar{J}^{b}_{-n}:
\end{equation}
which we have seen corresponds to the choice
$\al=-\bar{\al}$ and was argued not to be the correct choice for gravity. The terms $\pm 1$ in the
denominators are renormalization constants, and should be removed to obtain the classical
expression.
To clarify the difference of this choice compared to (\ref{wzwH}), insert (\ref{MJ1}) and
(\ref{MJ2})  into (\ref{wzwH}):
\begin{equation}\label{HKdS} H= \frac{1}{16} M,\end{equation} which shows that the Hamiltonian, as
expected, is proportional to the KdS$_3$ mass. The other possible choice of Hamiltonian
(\ref{MalStrHam}), involves a switch in sign in front of $\bar{L}_{0}$, and this
yields \begin{equation}\label{JKdS} H = -\frac{i}{16l}\J, \end{equation} i.e. the Hamiltonian in
this case is proportional to the spin rather than the mass of the KdS$_3$ geometry.

Recently a conjectural correspondence~\cite{Str:dSCFT} between quantum gravity on asymptotically de
Sitter space and conformal field theory on the conformal (spacelike) boundary has received much
attention. In the three-dimensional case analysis of asymptotic symmetries~\cite{BaldeBMin:AsdS,
Str:dSCFT} yields, in analogy with the classic work~\cite{Brown-Henneaux} concerning AdS$_3$, a
Virasoro algebra with central charge $c=3l/2G$. Naively inserted in the Cardy formula for the
asymptotic density of states in a modular invariant CFT it yields the Bekenstein-Hawking entropy for
the cosmological horizon~\cite{BouMalStr:01, Myung:01}. The CFT in question, however, is not
identified. Not surprisingly, our Virasoro algebra has effective central charge $c=6$, and it should
be interpreted as a CFT describing the gravitational sector of string theory in a KdS$_3$
background. Things are quite different in AdS$_3$ where we obtain a CFT description of the
gravitational sector of string theory with central charge $c\sim\frac{3l}{2G}$, i.e. the entropy is
obtained by calculating only gravitational degrees of freedom. Assuming validity of the dS/CFT
correspondence it thus seems that string theory on a dS$_3$ background couples matter and gravity
differently compared to AdS$_3$. This should perhaps not be considered surprising since there are
great difficulties in describing string theory in a dS background. Hopefully further development of
the dS/CFT correspondence will shed light on the nature of the differences between negative and
positive cosmological constant.


\newpage

\begin{thebibliography}{AA}
\bibitem{GibHawk:cosmhor} G.~Gibbons and S.~Hawking {\em Cosmological Event Horizons,
Thermodynamics, and Particle Creation}, Phys.~Rev.~{\bf D15} (1977) 2738-2751
\bibitem{Carlip:95} S. Carlip, {\em The Statistical Mechanics of the (2+1)-Dimensional Black Hole},
Phys.~Rev.~{\bf D51} (1995) 632-637, {\tt gr-qc/9409052}
\bibitem{BTZ} M. Banados, C. Teitelboim and J. Zanelli, {\em The Black Hole in Three-Dimensional
Spacetime}, Phys.~Rev.~Lett.~{\bf 69} (1992) 1849-1851, {\tt hep-th/9204099}
\bibitem{Carlip:98} S. Carlip, {\em What We Don't Know About BTZ Black Hole Entropy}, Class.~Quant.~
Grav.~{\bf 15} (1998) 3609-3625, {\tt hep-th/9806026}
\bibitem{FjHw:01} J.~Fjelstad and S.~Hwang {\em Sectors of Solutions in Three-Dimensional Gravity
and Black Holes}, Nucl.~Phys.~{\bf B628} (2002) 331-360, {\tt hep-th/0110235}
\bibitem{FjHw:99} J.~Fjelstad and S.~Hwang {\em Equivalence of Chern-Simons Gauge Theory and WZNW
Model Using a BRST Symmetry}, Phys.~Lett.~{\bf B466} (1999) 227-233, {\tt hep-th/9906123}
\bibitem{MaSu:00} T.~M{\aa}nsson and B.~Sundborg {\em Multi-Black Hole Sectors of AdS$_{3}$ Gravity}
Phys.~Rev.~{\bf D65} 024025, (2002) {\tt hep-th/0010083}
\bibitem{MaldStr:98} J.~Maldacena and A.~Strominger {\em Statistical Entropy of de Sitter Space},
JHEP~{\bf 9802}:014 (1998), {\tt gr-qc/9801096}
\bibitem{Str:dSCFT} A.~Strominger {\em The dS/CFT Correspondence}, JHEP~{\bf 0110}:034 (2001), {\tt
hep-th/0106113}
\bibitem{Park:KdS} M.-I.~Park {\em Entropy of Three-dimensional Kerr-De Sitter space}, Phys.~Lett.~{
\bf B440} (1998) 275-282, {\tt hep-th/9806119}
\bibitem{Witten:gravcs} E.~Witten {\em (2+1)-Dimensional Gravity as an Exactly Soluble System},
Nucl.~Phys.~{\bf B311} (1988) 46
\bibitem{Witten:complexCS} E.~Witten {\em Quantization of Chern-Simons Gauge Theory With Complex
Gauge Group}, Commun.~Math.~Phys.~{\bf 137} 29-66, (1991)
\bibitem{Bargmann:46} V.~Bargmann {\em Irreducible Unitary Representations of the Lorentz Group},
Ann.~Math.~{\bf 48} 3 (1947) 568-640
\bibitem{Naimark:linreps} M.A.~Naimark {\em Linear Representations of the Lorentz Group}, Pergamon
Press (1964)
\bibitem{Knapp:repth} A.~W.~Knapp {\em Representation Theory of Semisimple Groups, An Overview Based
On Examples}, Princeton University Press (1986)
\bibitem{Gawedzki:91} K.~Gawedzki {\em Noncompact WZW Conformal Field Theories}, Published in NATO
ASI: Cargese 1991:0247-274, {\tt hep-th/9110076}
\bibitem{Teschner:97_1} J.~Teschner {\em On Structure Constants and Fusion Rules in the
$SL(2,\CC)/SU(2)$ WZNW Model}, Nucl.~Phys.~{\bf B546} (1999) 390-422,{\tt hep-th/9712256}
\bibitem{Teschner:97_2} J.~Teschner {\em The Mini-Superspace Limit of the $SL(2,\CC)/SU(2)$ WZNW
Model}, Nucl.~Phys.~{\bf B546} (1999) 369-389, {\tt hep-th/9712258}
\bibitem{GivKutSeib:98} A.~Giveon, D.~Kutasov and N.~Seiberg {\em Comments on String Theory on AdS
$_{3}$}, Adv.~Theor.~Math.~Phys.~{\bf 2} (1998) 733-780 {\tt hep-th/9806194}
\bibitem{Teschner:99} J.~Teschner {\em Operator Product Expansion and Factorization in the $H_3^+$
WZNW Model}, Nucl.~Phys.~{\bf B571} (2000) 555-582, {\tt hep-th/9906215}
\bibitem{Brigham} N.~A.~Brigham, Proc.~Amer.~Math.~Soc. {\bf 1} (1950) 182
\bibitem{BaldeBMin:AsdS} V.~Balasubramanian, J.~de~Boer and D.~Minic {\em Mass, Entropy and
Holography in Asymptotically de Sitter Spaces}, (2001), {\tt hep-th/0110108}
\bibitem{Witten:01} E.~Witten {\em Quantum Gravity in de Sitter Space}, (2001) {\tt hep-th/0106109}
\bibitem{BouMalStr:01} R.~Bousso, A.~Maloney and A.~Strominger {\em Conformal Vacua and Entropy in
de Sitter Space}, (2001) {\tt hep-th/0112218}
\bibitem{Myung:01} Y.~S.~Myung {\em Entropy of the Three Dimensional Schwarzschild-de Sitter Black
Hole}, Mod.~Phys.~Lett.~{\bf A16} (2001) 2353, {\tt hep-th/0110123}
\bibitem{Brown-Henneaux} J. D. Brown and M. Henneaux, {\em Central Charges in the Canonical
Realization of Asymptotic Symmetries: an Example From Three-Dimensional Gravity}, Commun.~Math.~
Phys.~{\bf 104} (1986) 207-226
\end{thebibliography}
\end{document}